\documentclass[twocolumn,showpacs,preprintnumbers,amsmath,amssymb]{revtex4}
\usepackage{graphicx}
\usepackage{dcolumn}
\usepackage{bm}

\begin{document}


\title{The possibility of quasi-bound state formation\\ of $\eta$-meson with helium isotopes}

\author{V. A. Tryasuchev}
 \altaffiliation[Electronic address:]{ trs@npi.tpu.ru}
\author{A. V. Isaev}%
\affiliation{%
Tomsk Polytechnic University, Tomsk, Russia
}%



\begin{abstract}
The necessary conditions of quasi-bound state formation of
$\eta$-meson with isotopes $^{3}$He, $^{4}$He have been found within
the framework of optical potential model. These conditions have been
compared with the findings about helium nucleus densities and with
the available information about $\eta$N-scattering length. Thus, we
have conclude that within the framework of discussed model
$\eta-^{3}$He quasi-bound state formation is not possible, but
$\eta-^{4}$He quasi-bound state formation is possible with the great
probability.
\end{abstract}

\maketitle

The interaction of $\eta$-mesons with helium isotopes has been
considered in the frame of optical potential for the purpose of
$\eta-^{3}$He and $\eta-^{4}$He quasi-bound state formation. Let us
connect the optical potential distributions with the nuclear
densities of the discussed nuclei using the findings of
root-mean-square radii. For the description of low energy
$\eta$N-interaction let us use a well-known fact of resonance
domination $S_{11}$(1535) in the amplitude of this interaction at
such energies. In that case the optical potential of $\eta$-$A$
interaction of $U(r)$ takes the following form \cite{Er}:
\begin{equation}
    2\mu U(r) = -4\pi(1+\frac{m_{\eta}}{m_{N}})\rho (r)a_{0},
\end{equation}
where, $m_{\eta}$, $m_{N}$ are the meson and nucleon masses, $\mu$
is the reduced meson-nucleus mass, $a_{0}$ is the $\eta$N-scattering
length, $\rho(r)$ is the spherically symmetrical density of nucleon
in nuclei, which has been chosen in Fermi form:
\begin{equation}
    \rho (r)=\frac{\rho _{0}}{1+\exp (\frac{r-R_{c}}{a})}.
\end{equation}
Here $R_{c}$ is the half-density radius, $a$ is the thickness of
nucleus dif\mbox{}fusion surface layer, $\rho_{0}$ is the nucleon
density of nucleus in the center. For the nucleus with the nucleon
number $A$, two parameters in distribution (2) may can be fixed by
the conditions:
\begin{equation}
A = \int^{\infty}_{0}r^{2}\rho (r)dr, \quad \left\langle
r^{2}\right\rangle = \frac{1}{A}\int^{\infty}_{0}r^{4}\rho (r)dr,
\end{equation}
where $\left\langle r^{2}\right\rangle^{1/2}$  is the
root-mean-square radius ($r_{rms}$) of the nucleus. The knowledge of
rms radii of $^{3}$He, $^{4}$He nuclei \cite{El} leaves only one
free parameter of $U(r)$ radial distribution, which is called
``dif\mbox{}fuseness" and stand for $a/R_{c}$. The nucleus densities
depend on the dif\mbox{}fuseness parameters, as it may be seen in
figures 1, 2.

For the formation of quasi-bound state in the complex potential with
complex energy eigenvalue $E=-(\varepsilon+i\frac{\Gamma}{2})$ where
$\varepsilon$ is the binding energy, and $\Gamma$ is the level
width, the definite relation between absolute values of imaginary
and real parts of this potential is required, at which the bound
state is possible \cite{Tr,Si}. The calculated formation boundaries,
$\varepsilon\approx0$, of the discussed $\eta$-nuclei in the
dependence of imaginary potential part on the real one for
dif\mbox{}ferent nucleon distributions in nuclei $^{3}$He and
$^{4}$He are shown in figures 3, 4 in the complex plane of free
$\eta$N-scattering length. It is evident that at the nucleus
dif\mbox{}fuseness decrease for the quasi-bound state formation the
greater real potential part is required, that is, real parts of
$\eta$N-scattering length. At the dif\mbox{}fuseness increase, when
$a/R_{c}$ is over 0.25, the formation boundaries of $\eta$-nuclei
practically stop shifting to the left, that limits the dependence of
quasi-bound state formation on the nuclear density distribution. In
view of impossibility of complex $\eta$N-scattering length
experimental determination $a_{0}$ is found indirectly, that is,
they are model dependent and dif\mbox{}fer greatly from paper to
paper \cite{Ar,Gr,Ba,Ka,Ab,Sa,Wi,Ti} (see table 1 in paper
\cite{Ha}). The values of $\eta$N-scattering length getting into the
darkened areas in figures 3, 4 demonstrate the possibility of
$\eta$-nucleus formation at the attraction potential initiated by
such $a_{0}$. And on the contrary, if these values are left in the
white parts of figures 3, 4 the formation is not possible.

The bound states spectrum simulating results in $\eta-^{3}$He system
are contradictory \cite{Ha,Fi,Ke} and ambiguous \cite{Ch}. The same
situation presents in calculations of bound states in $\eta-^{4}$He
system \cite{Ha,Bu}. One may see in figure 3 that formation of
quasi-bound state $\eta-^{3}$He for the known data concerning length
of $\eta$N-scattering and root-mean-square radius of $^{3}$He
nucleus is impossible for any dif\mbox{}fuseness of $^{3}$He nucleus
density. If we use ``off-shell" length of $\eta$N-scattering in
potential (1) as authors of paper \cite{Ha} insist, the conclusion
will be the same because the real and imaginary parts of $a_{0}$
should, according to the cited paper, decrease proportionally. On
the contrary, the existence of quasi-bound state of $\eta-^{4}$He
within the limits of used model is possible and is almost
independent from the dif\mbox{}fuseness of $^{4}$He nucleus density
as it may be seen in figure 4 if $|Re(a_{0})| \geq 0.60$ fm.

\begin{figure}
\includegraphics{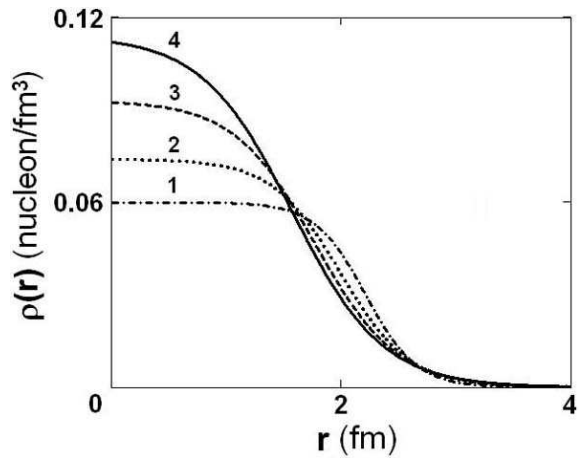}
\caption{\label{fig:epsart} The nucleon density distributions of
$^{3}$He nucleus for dif\mbox{}ferent dif\mbox{}fuseness parameter
values ${a}/R_{c}$ and fixed root-mean-square radii. The curve
parameters are given in table 1.}
\end{figure}
\begin{table}
\caption{\label{tab:table1}Curves for nucleus $^{3}$He}
\begin{ruledtabular}
\begin{tabular}{ccccc} $r_{rms}$ (fm) & $N^{0}_{-}$ & $a/R_{c}$ &
$R_{c}$ (fm) & $\rho_{0}$ (nucleon/fm$^{3}$)\\\hline
      & 1 & 0.10 & 2.210 & 0.060\\
1.9   & 2 & 0.15 & 1.991 & 0.074\\
      & 3 & 0.20 & 1.770 & 0.093\\
      & 4 & 0.25 & 1.571 & 0.114\\
\end{tabular}
\end{ruledtabular}
\end{table}

\begin{figure}
\includegraphics{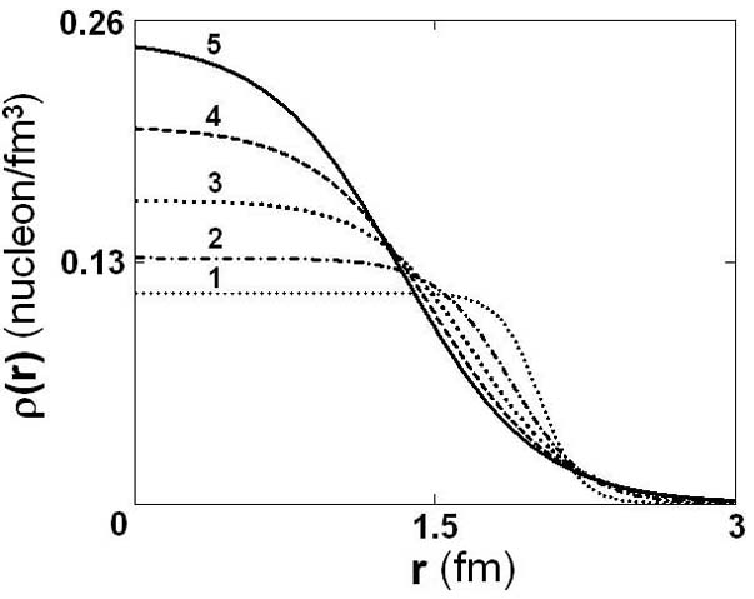}
\caption{\label{fig:epsart} The nucleon density distributions of
$^{4}$He nucleus for dif\mbox{}ferent dif\mbox{}fuseness parameter
values ${a}/R_{c}$ and fixed root-mean-square radii. The curve
parameters are given in table 2.}
\end{figure}
\begin{table}
\caption{\label{tab:table2}Curves for nucleus $^{4}$He}
\begin{ruledtabular}
\begin{tabular}{ccccc} $r_{rms}$ (fm) & $N^{0}_{-}$ & $a/R_{c}$ & $R_{c}$ (fm) & $\rho_{0}$
(nucleon/fm$^{3})$\\\hline
      & 1 & 0.05 & 2.021 & 0.113\\
      & 2 & 0.10 & 1.874 & 0.132\\
1.6   & 3 & 0.15 & 1.687 & 0.163\\
      & 4 & 0.20 & 1.499 & 0.203\\
      & 5 & 0.25 & 1.331 & 0.250\\
\end{tabular}
\end{ruledtabular}
\end{table}

\begin{figure*}
\includegraphics{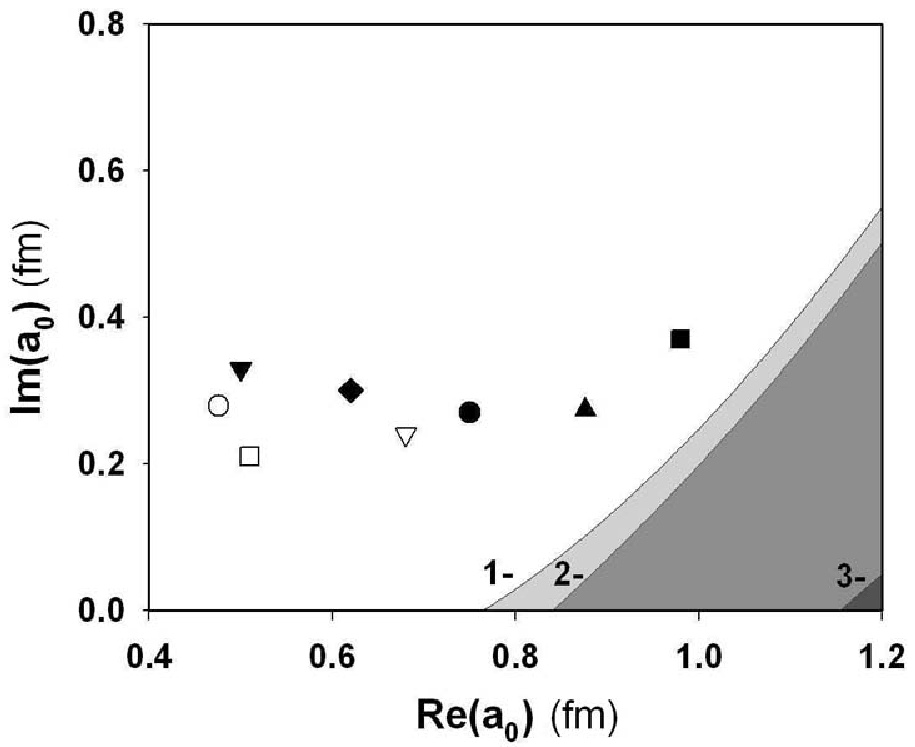}
\caption{\label{fig:wide}Curves are the boundaries of quasi-bound
states in system $\eta-^{3}$He. The darkened areas are the areas of
quasi-bound state formation of $\eta$-meson with $^{3}$He nucleus in
the complex plane of $\eta$N-scattering length for dif\mbox{}ferent
dif\mbox{}fuseness $a/R_{c}$ parameters: $1 - 0.25$; $2 - 0.15$; $3
- 0.1$. $\eta$N-scattering lengths have been taken from works:
\small $\blacksquare$ \normalsize $-$ $[5]$; \Large $\bullet$
\normalsize $-$ $[6]$; \normalsize $\blacktriangle$ $-$ $[7]$;
$\triangledown$ $-$ $[8]$; $\blacklozenge$ $-$ $[9]$; \small
$\square$ \normalsize $-$ $[10]$; $\blacktriangledown$ $-$ $[11]$;
\Large $\circ$ \normalsize $-$ $[12]$.}
\end{figure*}

\begin{figure*}
\includegraphics{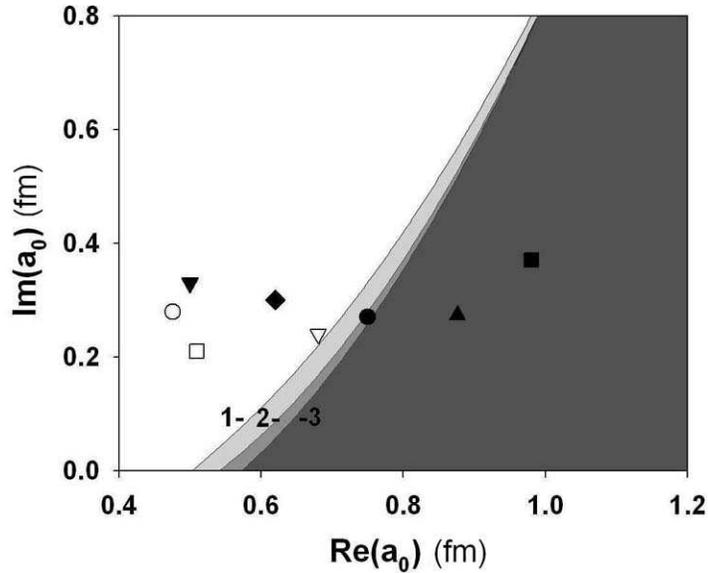}
\caption{\label{fig:wide}Curves are the boundaries of quasi-bound
states in system $\eta-^{4}$He. The darkened areas are the areas of
quasi-bound state formation of $\eta$-meson with $^{4}$He nucleus in
the complex plane of $\eta$N-scattering length for dif\mbox{}ferent
dif\mbox{}fuseness $a/R_{c}$ parameters: $1 - 0.25$; $2 - 0.15$; $3
- 0.05$. $\eta$N-scattering lengths have been taken from works:
\small $\blacksquare$ \normalsize $-$ $[5]$; \Large $\bullet$
\normalsize $-$ $[6]$; \normalsize $\blacktriangle$ $-$ $[7]$;
$\triangledown$ $-$ $[8]$; $\blacklozenge$ $-$ $[9]$; \small
$\square$ \normalsize $-$ $[10]$; $\blacktriangledown$ $-$ $[11]$;
\Large $\circ$ \normalsize $-$ $[12]$.}
\end{figure*}

The experimental result for both reactions $d + d$ $\longrightarrow$
$^{4}$He $+$ $\eta $, $d + p$ $\longrightarrow$ $^{3}$He $+$ $\eta $
near the thresholds point the features. The reaction's
cross-sections beginning at the thresholds do not increase with
increase of final particle's momenta \cite{Bu,Me}. This fact does
not contradict to the formation of quasi-bound states of
$\eta$-meson with these nuclei, but such features of cross-section
dependencies may be explained by the poles in the amplitudes of
$\eta^{3}$He- and $\eta^{4}$He-scattering, that are located directly
near the reaction's thresholds. Our conclusion concerning the
impossibility of formation of bound states in $\eta-^{3}$He system
in accord with calculation result of $\eta^{3}$He-interaction in
many-particle approach \cite{Fi} but contradicts to the conclusion
of theoretical papers \cite{Ha,Ke} in which the systems of
$\eta-^{3}$He in order to find bound states were considered using
different approximations. The result of the latest experiment from
COZY \cite{Kr} doesn't approve the possibility of $\eta$-nucleus of
$^{3}$He system formation. Another conclusion concerning a
possibility of bound state formation in $\eta-^{4}$He system does
not contradict to result of paper \cite{Ha} but in our case it is
possible at larger values of real parts of $\eta$N-scattering
length.

In the conclusion we may say that basing on the exact solution of
Schrodinger equation with optical potential (1) and Woods-Saxon
radial dependence that bound state $\eta-^{4}$He is possible, but
one of $\eta-^{3}$He is not. The features of virtual state of
$\eta-^{3}$He have been observed in the experiment \cite{Me}, while
it may be that experiment \cite{Bu} the features of bound state of
$\eta-^{4}$He have been observed.

Authors thank for help in the work L. Sukhikh.

\newpage 
\bibliography{izotops}

\end{document}